FERMILAB-CONF-18-017-AD# PIP-II INJECTOR TEST WARM FRONT END: COMMISSIONING UPDATE*

L. Prost[†], R. Andrews, C. Baffes, J.-P. Carneiro, B. Chase, A. Chen, E. Cullerton, P. F. Derwent,
J. P. Edelen, J. Einstein-Curtis, D. Frolov, B. Hanna, D. Peterson, G. Saewert, A. Saini,
V. Scarpine, A. Shemyakin, J. Steimel, D. Sun, A. Warner, Fermilab, Batavia, IL 60510, USA
C. Richard, Michigan State University, East Lansing, MI, USA
V.L.S. Sista, Bhabha Atomic Research Centre (BARC), Mumbai, India*Abstract*

The Warm Front End (WFE) of the Proton Improvement Plan II Injector Test [1] at Fermilab has been constructed to its full length. It includes a 15-mA DC, 30-keV H$^-$ ion source, a 2 m-long Low Energy Beam Transport (LEBT) with a switching dipole magnet, a 2.1 MeV CW RFQ, followed by a Medium Energy Beam Transport (MEBT) with various diagnostics and a dump. This report presents the commissioning status, focusing on beam measurements in the MEBT. In particular, a beam with the parameters required for injection into the Booster (5 mA, 0.55 ms macro-pulse at 20 Hz) was transported through the WFE.
## PIP2IT WARM FRONT END

The PIP2IT WFE (Fig. 1) as described in Ref. [2] has been installed in its nearly final configuration. Details of the H$^-$ ion source, LEBT and RFQ can be found in Refs. [3-6], as well as their performance. The combination of the ion source and LEBT can deliver up to 10 mA at 30 keV to the RFQ with pulse lengths ranging from 1 μs to 16 ms at up to 60 Hz, or a completely dc beam. An atypical transport scheme was devised in the LEBT [7], with which changes of the beam properties throughout a pulse due to neutralization are eliminated, hence facilitating tuning of the beam line at any pulse length (including dc). Following the RFQ is a long MEBT [8], which provides transverse and longitudinal focusing to match the 2.1-MeV beam into the Half-Wave Resonator (HWR) cryomodule. As the latter is not yet installed, the beam line currently ends with a high-power dump capable of dissipating 15-20 kW, depending on the beam size.

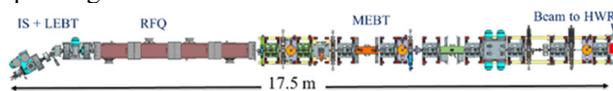

Figure 1: PIP2IT warm front end (top view).

The MEBT (Fig. 2) can be described in terms of the succession of nine "sections" (650-mm long flange-to-flange for sections #1 through #7), delimited by transverse focusing assemblies, except for the last section that includes several diagnostics and the beam dump. Transverse focusing is provided by quadrupoles grouped into doublets (2) and triplets (7) [9]; each group includes a Beam Position Monitor (BPM) button, whose capacitive pickup is bolted to the poles of one of the quadrupoles. Longitudinal focusing is provided by 3 bunching cavities that can operate in either pulsed or Continuous Wave (CW) mode. The chopping system consists of two travelling-wave kickers separated by 180º transverse phase advance and an absorber at a 90º phase advance from the last kicker. The absorber is followed by a Differential Pumping Insert (DPI), which 200 mm (L) × 10 mm (ID) beam pipe reduces the flux of gas released from the bombardment of the absorber with H$^-$ ions into the cryomodules downstream.

Movable scrapers [10] are used to measure the beam size (low power operation), as protection against errant beam or halo (most importantly for high power operation) and to intercept one of two trajectories when characterizing the kickers performance. In the present beam line, there are 4 sets of 4 scrapers (each set consists of a bottom, top, right and left scraper) plus a temporary set of two scrapers (a.k.a. F-scraper, top and right) located just downstream of the prototype absorber but before the DPI.

Current transformers are located at the beginning and end of the MEBT. An emittance scanner and Fast Faraday Cup (FFC) (moved to various locations) were used to characterize the beam emittance. A Resistive Wall Current Monitor (RWCM) completes the set of diagnostics available.

## MEBT COMMISSIONING

Reports on commissioning activities for the Ion Source, LEBT and RFQ can be found in Refs. [4,5,11,12] and therein. This section focuses on the MEBT.

### MEBT Chopping System

The chopping system synchronously uses two kickers to provide a 6-$\sigma$ separation at the absorber (where $\sigma$ is the beam rms vertical width at that location) between the bunches intended to be removed and those making their way past it. Thus, two kicker prototypes of different designs, termed "50-Ohm" and "200-Ohm" [13] in accordance with their respective impedance, have been installed in the beam line for evaluation.

To characterize them individually, dipole correctors, both before and after the kicker being tested, are set such that the beam is transported to the dump with low losses whether the kicker is on or off (i.e. 2 separate trajectories simultaneously). Note that this was only possible before installation of the DPI, which creates a severe aperture restriction, incompatible with this manipulation.

___________________________________________
* This manuscript has been authored by Fermi Research Alliance, LLC under Contract No. DE-AC02-07CH11359 with the U.S. Department of Energy, Office of Science, Office of High Energy Physics.
[†] lprost@fnal.gov

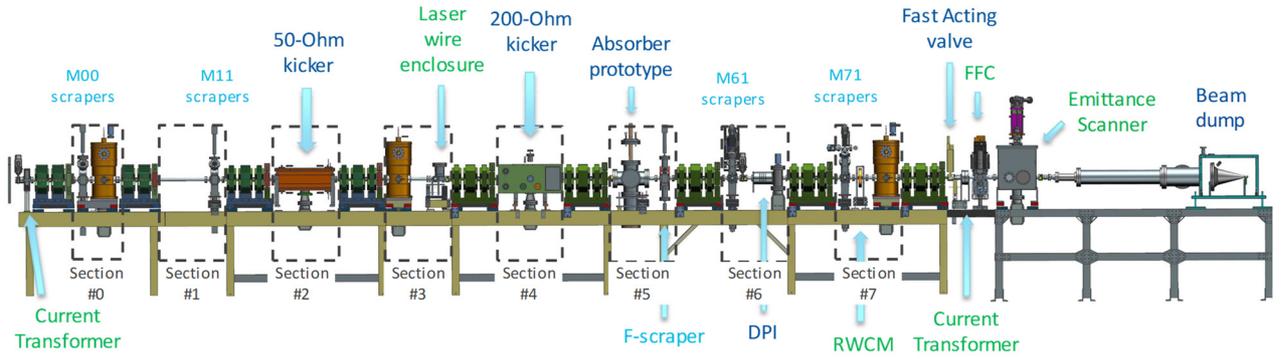

Figure 2: Medium Energy Beam Transport line (side view).

This configuration permitted demonstrating operation of the 200-Ohm kicker for an extended period of time at 5 mA × 0.55 ms × 2.1 MeV × 20 Hz = 115 W, a.k.a. "CDR MEBT parameters" [14] for injection into the Booster. While similar trajectories were devised for testing the 50-Ohm kicker, no long run was attempted.

To measure the deflection strength of the kickers, scraper profiles of the beam downstream are recorded (Fig. 3) when every other bunch is deflected (81.25 MHz waveform). As the scraper moves into the beam, it first intercepts one of the trajectories, then the second. The dump current is plotted against the scraper position and, if the two trajectories are well separated, shows two 'steps' (Fig. 3a). On the other hand, if the deflection is small, the two 'steps' start merging as shown on Figure 3b.

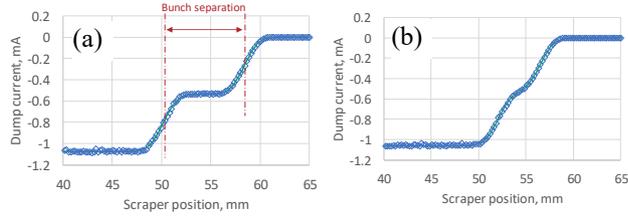

Figure 3: F-scraper profiles for different delays between the phase of the beam and the kicker's. (a) Optimum delay; (b) Shifted by 1.73 ns w.r.t. the optimum delay. The beam current after scraping is 1.1 mA (5 mA initially).

Initially, phasing of the kickers w.r.t. the beam is done by recording the signal of a single BPM plate downstream of the kicker on an oscilloscope and maximizing the peak amplitude difference between successive bunches by adjusting the relative delay.

Quantitatively, the scraper profiles are fitted with a function corresponding to the sum of 2 Gaussian 1D profiles, where the fitting parameters are the mean value of each Gaussian and $\sigma$, which is assumed to be the same for bunches belonging to either trajectory. The difference between the 2 mean values is the trajectory or bunch separation. From this, both the optimum phase of the kicker w.r.t. the beam and the deflection angle at the kicker exit can be inferred. Note that because a single kicker only provides half of the design separation for the nominal beam, for the 2 mean values to be well separated at nominal voltage, the beam is scraped off in the direction of the kick (vertical) with the first scraping station (resulting in a so-called flat beam).

In this arrangement, both designs met the deflection specification of >7 mrad between passing and removed bunches, which for the 200-Ohm kicker corresponds to a nominal voltage difference of 1kV between the 2 opposite electrodes.

The higher impedance of the 200-Ohm kicker allows using state-of-the-art fast switches developed at Fermilab [15], and, in turn, demonstrating arbitrary bunch selection without the need to invest in expensive power amplifiers that would be required to fully test a 50-Ohm system. The present version of the switch-driver can deliver 0.6 ms bursts of arbitrary pulses with 4 ns rise/fall time, average switching frequency of up to 45 MHz inside the burst and 20 Hz bursts repetition rate as required for Booster injection. This is illustrated in Figure 4 with the waveform of the RWCM located downstream of the 200-Ohm kicker. A flat beam was used, and the F-scraper is partially inserted into the beam pipe such that deflected bunches get intercepted by the scraper paddle.

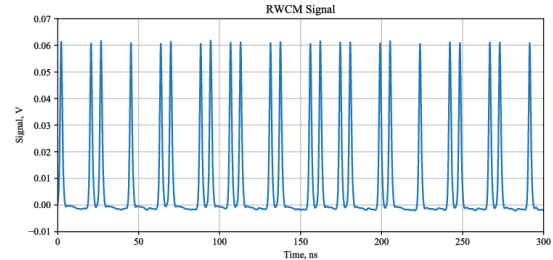

Figure 4: RWCM waveform showing part of the bunch pattern tailored for Booster injection created by the 200-Ohm kicker.

Finally, both kickers were used together synchronously using 81.25 MHz waveforms. In that case, one of the 2 trajectories created by the 50-Ohm kicker was further deflected by the 200-Ohm kicker, while losses to the kicker masks, electrically isolated metal sheets which protect their electrodes, were kept to <20 μA altogether. This configuration is illustrated on Figure 5, which shows the ±3$\sigma$ vertical envelopes from Tracewin [16] simulations for both the passing beam (pink trace) and the beam being deflected onto the F-scraper (green trace).

Note that for the simulations shown in Figure 5, the focusing settings are those used experimentally and were not completely optimized.

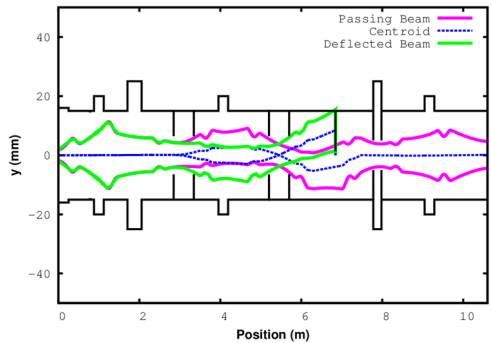

Figure 5: $3\sigma$ vertical envelope simulated with Tracewin showing both the passing and the deflected bunches in the configuration with both kickers used in sync.

*Beam Properties*

While operation of the Warm Front End was demonstrated for a beam with "CDR parameters" (i.e. 1.1% duty factor), most beam properties measurements are carried out with short pulses (10 µs). Scrapers, the Fast Faraday Cup (FFC) and the emittance scanner are all intercepting devices that cannot sustain large beam power deposition.

Transverse beam size measurements with the scrapers and emittance scanner in conjunction with magnetic measurements of the focusing elements indicate that the MEBT optics are understood and predictable at the 10% level [17]. As mentioned previously, this knowledge allows transporting the beam through the MEBT with <2% loss while going through the tight apertures of the 2 kickers and, even more so, of the DPI. For the nominal beam current of 5 mA (out of the RFQ), with the kickers off, the transverse emittance measured at the exit of the MEBT is within the specifications (< 0.22 mm mrad, rms, normalized) and nearly the same as when it was measured at the exit of the RFQ (earlier configuration when there were only 2 quadrupole doublets followed by the emittance scanner [11]). In addition, when every other bunch is either kicked out or passes through, the emittance of the 'passing' bunches show <10% increase.

Measurements of the rms longitudinal emittance have been performed with the Fast Faraday Cup (FFC) when it was located just upstream of the 3rd bunching cavity (Section 7, Fig. 2). A first estimate gives a value of 1.06 eV µs (i.e. 0.34 mm mrad normalized) at 5 mA for a 10 µs pulse, ~10% higher than expectations.

## HIGH POWER OPERATION

As mentioned in Ref. [11] operation of the RFQ with CW RF was put on hold after the RFQ couplers developed vacuum leaks. While pulsed RF operation continued with spare couplers, new units in which the ceramic windows that isolate the vacuum from atmosphere are not brazed to the coupler chamber but sealed with O-rings instead, have been designed and installed recently. As a result, CW RF operation of the RFQ is now standard, which allows exploring operation at higher beam power.

Nevertheless, even for pulsed beam operation, uncontrolled losses in the MEBT can damage the beam line components when the beam pulse > 0.1 ms. Therefore, a robust Machine Protection System (MPS) is needed.

With low energy protons/H-, traditional beam loss monitors that rely on ionizing radiation produced by the beam when it interacts with the vacuum chamber is not possible. Thus, the PIP2IT MPS is based on comparing beam current monitor readings along the beam line (similarly to SNS's MPS scheme [18]) and direct measurement of losses onto various electrodes.

For beam loss measurements, there are 4 capacitive Ring Pick-Ups (RPU), similar to button BPMs, dedicated to the Machine Protection System (MPS): one pair measures the beam lost upstream of the absorber and the second, downstream of the absorber. Because RPUs are not direct current measuring devices, they are calibrated in-situ with short beam pulses using AC current transformers for reference. One advantage of the RPU is that the signal output does not depend on the pulse length including purely CW.

In addition to RPUs, limits on beam loss are imposed by measuring the current of all scraper paddles, as well as the kicker masks and the DPI, which is also electrically isolated. Presently, noise in the RPUs limit the protection scheme to the ~3% level.

With the RFQ operating in CW mode, increasing the beam power transported through the RFQ, MEBT and to the dump, is carried out by increasing the macro-pulse length that the LEBT chopper delivers. So far, a beam of up to 1 kW was successfully transported to the dump over more than 24 hours with 96% uptime. Current efforts aim at increasing the power delivered to the dump to 10 kW.

## PLANS

With both kickers performing well with beam, the kicker technology for PIP-II was chosen to be the 200-Ohm version for which the driver has also be proven to be capable of generating arbitrary bunch patterns. Two new (production) kickers will be fabricated, with minor design modifications based on the experience with the prototype.

The MEBT absorber rated for 21 kW CW has been designed and is being fabricated. It is expected to be tested by the end of the year.

Then operation will be stopped for installation of the cryogenics distribution system and cryomodules. During the shutdown, the final version of the kickers, absorber, fast vacuum protection of cryomodules, and particle-free components of the MEBT will be installed. Beam operation will resume in 2020, with the warm front end initially delivering a pulsed beam into the cryomodules. At that stage, focus will be brought to vacuum management near the SRF structure for normal operation with "CDR" beam parameters.

## ACKNOWLEDGMENT

The authors are thankful to the many people who contributed to building PIP2IT and helped with its operation, including but not limited to K. Carlson,

M. Coburn, J. Czajkowski, N. Eddy, B. Fellenz, D. Franck, M. Ibrahim, T. Hamerla, M. Hassan, S. Kazakov, K. Kendziora, S. Khole, M. Kucera, D. Lambert, W. Mueller, R. Neswold, D. Nicklaus, A. Saewert, D. Sharma, J. Simmons, T. Zuchnik.